\documentclass[epj]{svjour}
\usepackage{graphics,amsmath}

\usepackage{graphics}
\begin{document}
\title{Large-scale low-energy excitations in 3-d spin glasses}
\subtitle
{Replica symmetry breaking and characterization in position space}
\author{J. Houdayer\inst{1} \and F. Krzakala\inst{2} \and O. C. Martin\inst{2}}
\offprints{krzakala@ipno.in2p3.fr} 
\institute{Institut f\"ur Physik, Mainz, Germany
\and LPTMS, Univ. Paris-Sud, Orsay, France}
\date{\today}

\abstract{
We numerically extract large-scale excitations above the ground state in 
the 3-dimensional Edwards-Anderson spin glass with Gaussian
couplings. We
find that associated energies are $O(1)$, in agreement with 
the mean field picture. Of further interest are
the position-space properties of these excitations.
First, our study of their {\it topological} properties show that
the majority of the large-scale excitations 
are sponge-like. Second, when probing their {\it geometrical}
properties, we find that the excitations coarsen when the system size
is increased. We conclude that
either finite size effects are very large even
when the spin overlap $q$ is close to zero, or
the mean field picture of homogeneous excitations has to be modified.
}
\maketitle
\section{Introduction}
\label{intro}
The physics of spin glasses~\cite{Young98} is an old 
yet still very active
subject of study. Much progress has been made in understanding
the nature of the low temperature phase, but some fundamental
issues remain open. Among these is whether the free-energy
landscape consists of multiple valleys all contributing to the
partition function in the thermodynamic limit. This is
what happens in mean field theory~\cite{MezardParisi87b}, 
and a growing consensus is that it
also arises in finite dimensions. Perhaps the strongest numerical 
evidence in favor of this comes from 
the spin overlap probability distribution $P(q)$
that seems to be non-trivial (see~\cite{MarinariParisi99b} 
for a review). 

Several other striking properties
of mean field may also hold in finite dimensions;
the purpose of our work is to test this
by numerically characterizing the valley states in the energy landscape
of the 3-dimensional Edwards-Anderson~\cite{EdwardsAnderson75} (EA)
spin glass model. In the mean field picture, one expects
some system-size excitations (configurations obtained from the
ground state by flipping a finite fraction of all the spins)
to have excess energies of $O(1)$. Furthermore,
these excitations should
(i) be sponge-like with non-zero surface to volume ratios;
and (ii), be homogeneous on scales larger than the lattice spacing.

We first describe how
we find excitations in Section~\ref{sect_excitations} and then
discuss in Section~\ref{sect_scenarios}
some possible scenarios for how replica symmetry breaking may arise
in position space (unfortunately sometimes called 
real space).
In Section~\ref{sect_topology} we investigate topological
properties of system-size excitations. Previous 
work~\cite{KrzakalaMartin00}
gave evidence that the $3-d$ EA model 
with free boundary conditions
had system-size
excitations whose energies were $O(1)$. In the present work, we use
periodic boundary conditions and consider the excitations' 
windings around the lattice to classify them topologically. We conclude
again that it is possible to flip a finite fraction of all the
spins at an $O(1)$ energy cost. Furthermore,
most of these excitations are sponge-like, and our data are compatible
with the possibility that
{\it all} system-size excitations are sponge-like in the
thermodynamic limit. Then in Section~\ref{sect_geometry} we
try to characterize the geometrical
properties of these excitations. We measure 
their surface area and correlation functions, and
then probe the distribution of their hole sizes.
The data is most naturally interpreted
as indicating that there is an intrinsic length scale that
grows with the lattice size; if this scale grows indefinitely, then 
surface to volume ratios of the excitations 
go to zero~\cite{KrzakalaMartin00,PalassiniYoung00a} 
asymptotically. Beyond this scale
(which is much smaller than the lattice size), excitations may be
homogeneous, in agreement with the mean field picture.
Given the small range in sizes that we can treat, it is also possible,
though less likely, that the growth of
this length scale is a finite size effect which saturates for
larger sizes; if this is the case, the mean field picture
holds and excitations are
spongy beyond a few lattice spacings; whether they are homogeneous
or lumpy is less clear as we find evidence for heterogeneities 
on scales much larger than the lattice spacing.

\section{System-size excitations}
\label{sect_excitations}
\subsection{Valleys}
\label{subsect_valleys}
Consider a general Ising spin glass model with pairwise couplings
\begin{equation}
H = - \sum_{i<j} J_{ij} S_i S_j
\end{equation}
Let $N$ be the total number of spins.
We would like to characterize the energy landscape of this system through the
statistical properties of its valleys. In the infinite volume case,
one can define a valley via the configuration at its bottom;
the bottom of a valley can be defined as a minimum
of the energy under all {\it finite} number of spin 
flips. Unfortunately, this simple
definition has to be modified in finite volume; a simple 
yet heuristic modification consists in 
imposing the energy to be a minimum under any number of spin
flips fewer than $v N$.
$v \in \lbrack 0,1/2 \rbrack$ is a distance parameter (in fact it
is the Hamming distance divided by $N$)
related to the usual spin overlap $q$ by $q = 1 - 2 v$.  Because
of the global spin flip symmetry of the Hamiltonian, 
hereafter we will force $q \ge 0$. 

The influence of $v$ on the definition of the 
valleys is important as all valley
bottoms by construction will differ by at least $v N$ spin flips.
Suppose one had a 1-step replica symmetry breaking with a spin
overlap probability distribution
\begin{equation}
P(q) = \alpha \delta(q) + (1-\alpha) \delta(q-q_1)
\end{equation}
If $q_1 \to 1$ as the temperature goes to zero,
all choices of $v$ satisfying
$v < 1/2$ would recover the thermodynamically
relevant valleys as $N \to \infty$.
However, in the EA model, one expects the replica
symmetry breaking to be continuous and valleys with 
overlaps arbitrarily close to 1 should arise in the zero
temperature limit. Then any fixed choice of $v$ will eliminate
some of the thermodynamically relevant valleys. Since this effect
decreases as $v$ does, one could simply
choose a small value of $v$ 
and determine empirically
the extent of the robustness in $v$ of the analysis.

However from a numerical point of view,
the main difficulty lies in {\it finding}
low energy states rather than in classifying them. In fact, it is
at present nearly impossible to find the minimum number
of spins to flip before the energy of a configuration is lowered.
Thus we cannot even tackle the task of finding the best valley
lying within a reduced distance $v$ and $(v + \Delta v)$
from the ground state. Because of this 
limitation, we have chosen a more hands-on
approach where we just {\it sample} low-lying excitations. Given
the way we find these configurations, we expect that nearly always
they will in fact be bottoms of valleys.

\subsection{Low energy excitations}
\label{subsect_procedure}
From an algorithmic point of view, the ground state configuration
${\cal C}^0$ of a spin glass 
can be determined with a high level of confidence as long as
$N$ is not too large. In this work, we have used a 
genetic renormalization 
algorithm~\cite{HoudayerMartin99b} which
allows us to go to lattices up to $12^3$.
Now for extracting valleys or excited states, 
it is necessary to first compute ${\cal C}^0$ and then
to introduce a ${\cal C}^0$-dependent constraint or
perturbation to the Hamiltonian
that will lead to a modified
ground state labeled ${\cal C}^1$. 

We have followed the constraint-based
technique proposed in~\cite{KrzakalaMartin00}:
given the ground state, we flip the relative orientation of
$2$ selected spins and compute the new ground state with this constraint.
A large scale excitation is obtained only if its energy is smaller
than that of all the droplets surrounding either of the two selected
spins. In practice, we repeat this multiple times by randomizing
the choice of the two selected spins, obtaining a collection
of ${\cal C}^1$ excitations.
Note that this method gives 
no control over the number $V$ of spins that will flip when 
going from ${\cal C}^0$ to ${\cal C}^1$.

For each instance (set of $J_{ij}$), our numerical simulation then leads
to ${\cal C}^0$ (the ground state), and a number of excitations.
From this set, one can extract the
configurations that satisfy any desired constraint. For instance,
one may impose $V/N \ge v_{min}$ when looking for excitations
of size growing linearly with $N$
and study the energy and position space
properties of these excitations.

\subsection{Two scales for finite energy excitations}
\label{subsect_two_scales}
Hereafter, we consider only the three-dimensional
Edwards-Anderson model with nearest-neighbor interactions. 
The $J_{ij}$ are i.i.d. Gaussian random variables 
of zero mean and unit variance,
and we work with $L \times L \times L$ cubic lattices ($L^3=N$)
and periodic boundary conditions.

Our procedure for sampling excitations can extract low energy
droplets, but within the 
mean field picture one expects to also find large-scale
excitations whose size grows {\it linearly} with $N$. 
We can test this two-scale 
picture by considering the probability distribution
of $V$, the number of spins flipped when going from
the ground state to the excited state. Suppose that the 
only excitations having $O(1)$ energy are the
droplets ($V$ finite as $L \to \infty$) and system-size excitations
($v=V/N$ finite as $L \to \infty$). Then excitations where
$V \to \infty$ but $v \to 0$ as $L \to \infty$
will have a probability tending towards $0$.
Mathematically, this means that our extraction
of excitations leads to a probability distribution for
$V$ of the form
\begin{equation}
\label{eq_localGlobal}
P_L(V) \underset{L \to \infty}{\longrightarrow}
\alpha P_l(V) + (1-\alpha) P_g(v)
\end{equation}
where $P_l$ and $P_g$ are the normalized probability distributions
associated with droplet (local) and system-size
(global) excitations. 

We have measured the 
frequencies of ``events'' with $V$ for instance in the range
$\lbrack L^2, 2L^2\rbrack$ as a function of $L$.
(We say that we have an event in our data set whenever an instance
and our sampling procedure lead to an excitation satisfying
chosen criteria. Here and in the rest of our study, we 
generated three excitations for each instance; we have
$10 000$ instances at each value of $L \le 10$, $8 000$ at 
$L=11$, and $5 000$ at $L=12$.) The frequencies decrease with
$L$ and quite reasonably can be extrapolated to zero. As an example,
for the interval given above,
we find the frequencies
$0.520$ ($L=4$), $0.397$ ($L=5$), $0.321$ ($L=6$), $0.256$ ($L=7$),
$0.232$ ($L=8$), $0.208$ ($L=9$), $0.187$ ($L=10$), $0.170$ ($L=11$),
and $0.162$ ($L=12$).
Note that by construction, one cannot get excitations
whose energies are larger than those of droplets, so all states found
in our sampling necessarily have $O(1)$ energies. 
This analysis then confirms the two-scale form given in
equation (\ref{eq_localGlobal}).
 
\section{Spin glass scenarios in finite dimensions}
\label{sect_scenarios}
\subsection{Length scales for self-averaging}
\label{subsect_self_averaging}
Consider first the {\it energy} of system-size
excitations. In the mean field picture (as 
motivated by the properties of the
SK model) we expect to be able to find a
${\cal C}^1$ (for any choice of $v_{min}$)
with a finite probability, and that its energy will be
$O(1)$, i.e., finite when $N \to \infty$. On the contrary,
in the droplet~\cite{FisherHuse88} and scaling~\cite{BrayMoore86}
pictures, the characteristic energy of large scale excitations
grows as a power of $L$. In~\cite{HoudayerMartin00b} it was argued
that one should distinguish the energy scaling of large-scale 
excitations from that of droplets in an infinite system; it is then
necessary to introduce two exponents,
$\theta_g$ ($g$ for global) and $\theta_l$ ($l$ for local); for instance
the lowest system-size excitations have energies 
scaling as $L^{\theta_g}$.
Section~\ref{subsect_sponges} will lead us to the conclusion that
$\theta_g \approx 0$. Extrapolating to finite temperature
as is now commonly done~\cite{MarinariParisi00b},
this indicates that replica symmetry is broken in the three-dimensional
Edwards-Anderson model, i.e., the overlap
probability distribution $P(q)$ is non-trivial in agreement with many
previous studies (see~\cite{MarinariParisi99b} for a review).

Consider now the
{\it position-space} characteristics of these large scale excitations. 
We are concerned here with the properties of the
(connected) cluster of spins that
are flipped when comparing to the ground state; we shall use the term
excitation when referring to this cluster in addition to the
configuration ${\cal C}^1$ itself. By definition, the sizes of these
clusters are $O(N)$ (those are the events we focus upon).
Of major interest is to understand whether these 
excitations are topologically non-trivial or not, homogeneous
or inhomogeneous, fractal, multi-fractal, etc... To classify the
different possibilities, we ask how the properties of the clusters
depend on the scale of observation and on $L$, the size of the
lattice. Clearly there are many observables that can be considered; to
stay as simple as possible this discussion will focus on the cluster's density
or equivalently on its local overlap as a function of the scale
of observation.

Consider an $M \times M \times M$ window or box and 
let $q_M$ be the overlap between ${\cal C}^0$ and ${\cal C}^1$ 
restricted to this box. (The cluster's density in the
box is just $(1-q_M)/2$.) Obviously $q_M$ fluctuates no matter how
large $M$ is as 
the global overlap $q$ itself is not self-averaging.
But if we fix $q$, at what scale in $M$ do the fluctuations in 
$q_M$ disappear, or equivalently, 
what is the typical fluctuation of $q_M$ as a function of 
the observation scale $M$? Let
$P_{M,L}(q_M)$ be the 
{\it disorder-averaged} distribution of $q_M$; its mean is $q$
and its variance will decrease to zero as $M$ approaches $L$.
We can distinguish different scenarios 
according to the behavior of this distribution.
We can expect and will assume that we have {\it pointwise} convergence
in the infinite volume, i.e,. that at fixed $M$
\begin{equation}
\label{eq_pointwise}
P_{M,L}(q_M) \underset{L \to \infty}{\longrightarrow} P_M(q_M)
\end{equation}
(Note that the global overlap between the two configurations must
be fixed, and so this distribution depends implicitly on $q$.)
It is possible that in fact this convergence will be
{\it uniform} in the mathematical sense, that is one converges
to the same function even if $M$ varies with $L$.
We then say that the limit is ``regular'', leading to
``regular'' scenarios. The other scenarios will be non-regular,
i.e., the convergence will be non-uniform. The key property that
distinguishes these scenarios is whether or not $q_M$ becomes
self-averaging at a fixed ($L$-independent) scale.
In the regular scenarios, the fluctuations in $q_M$ disappear
as $M$ grows, so that the scale for self-averaging is a few
(or many) times the lattice spacing. In the non-regular scenarios, 
$P_M(q_M)$ will not tend towards a delta function as
$M \to \infty$; fluctuations will go to zero only when $M$ is sufficiently
big compared to an $L$-dependent scale.

In physical terms, the regular scenarios have no hidden scale;
as soon as one takes large enough windows, one has convergence
of $q_M$ towards $q$ in the probabilistic sense. As a consequence,
once the global overlap $q$ has been fixed, there are essentially
no significant fluctuations in density down to a few lattice spacings,
$P_{M,L}$ being very peaked about its mean. $q_M$ 
is then self-averaging when $q$ is fixed
as soon as $M$ becomes large compared
to $1$. On the contrary, in the non-regular scenarios, 
$P_{M,L}$ does not become peaked about its mean.
For $q_M$ to become self-averaging, one needs to go out 
to an $L$-dependent scale, for instance 
$M \approx L^{\gamma}$. A simple system that realizes this scenario 
is a ferro-magnet with anti-periodic boundary conditions; then $\gamma = 1$,
and for scales 
$M \ll L$, $P_{M,L}$ has a two peak structure; 
$q_M$ is not self-averaging on any scale smaller than $L$. 

Our presentation has concentrated on the 
behavior of {\it density} fluctuations
on different scales, but it generalizes in a straightforward manner
to other observables (surface of the clusters, number 
of handles, etc...). Since
different observables may become self-averaging on different
scales, it is best to keep in mind that
the complete picture may be more complicated than
that obtained by looking at density fluctuations alone.

\subsection{Two scenarios with uniform convergence}
\label{subsect_uniform}
With {\it uniform} convergence, the limit $L \to \infty$
in equation~\ref{eq_pointwise} need not be restricted to 
$M$ fixed; any $M$ (dependent or not on $L$) will lead to
the value given by the pointwise limit. 
We can say that there is a smooth or regular 
infinite volume limit on all scales simultaneously.

Our classification of
regular scenarios is based on the way 
$P_M$ converges towards a (single!) delta function as $M \to \infty$.
Zooming into the region around its peak, one may expect
$P_M$ to have a limiting shape if we rescale the
$x$ axis so that the full-width half-max stays constant.
If the limiting shape, $P^*$, decays quickly so all moments
exist, then fluctuations much larger than typical ones
will be very rare and we will say that the system is homogeneous.
If, on the contrary, $P^*$ is a 
{\it broad} distribution, then rare events
can dominate the mean square of density fluctuations,
and we will say that the system is heterogeneous.
Let us give physical pictures of these two cases.

\subsubsection*{Regular homogeneous sponges}
Consider first the regular {\it homogeneous} scenario. A well known
system having this behavior is the Ising model near 
its Curie point. Long wave-length
fluctuations occur and decrease in intensity with wave-length.
The dependence of this intensity on the scale ($M$) is not the point
here, rather we focus on the {\it distribution} of the fluctuations
at given $M$. In this physical system,
fluctuation intensities much larger than 
the typical ones are very rare, the probability distribution
$P^*$ decays faster than any power.
For our spin glass problem, we will have the
regular homogeneous scenario if the density fluctuations of
the clusters have this same property as long as we keep
the global overlap $q$ fixed. In a pictorial language, the 
clusters associated with an
excitation will be rather random at scales larger than the
lattice spacing, and can be thought of as homogeneous
sponges characterized mainly by their density $v$ and
their effective mesh spacing $\ell_c$. The terminology ``sponge''
comes from the fact that such objects are full of handles
and are bicontinuous: they and their complement are
connected, and the local density beyond the scale $\ell_c$
is nearly uniform. This length $\ell_c$, called the cohesive
length of the sponge~\cite{HoudayerMartin00b}, is just a few lattice
spacings. (Since the convergence is uniform, the lattice spacing
is the only scale, $\ell_c$ cannot diverge with $L$ and it
is unnatural for it to be many lattice spacings.)

\subsubsection*{Regular inhomogeneous sponges}
\label{subsubsect_inhomogenous}
A different scenario is obtained
if we give up the homogeneity hypothesis.
We still maintain uniform convergence of observables towards their
large $L$ limit, but now we assume that the density fluctuations 
arise not via low intensity modes but via holes in the sponges.
To use the picture of Villain, we have something like a swiss cheese,
except that the starting point is a sponge rather than the whole 
lattice, and from this we take out (or put in) blocks of contiguous
spins. The resulting object is lumpy on all finite scales and
the distribution of hole sizes and thus
$P^*$ has long tails. But
as long as large holes
are sufficiently rare, we can maintain the uniform convergence
so that this lumpy system has no characteristic
scale other than the lattice spacing.

\subsubsection*{The mean field picture}
In any uniform convergence or regular scenarios,
a finite fraction
of the sites in the cluster are at a finite distance from the
cluster's complement; thus the surface area of the cluster grows as
$L^3$, and the link overlap $q_L$ cannot go to 
$1$ as $L \to \infty$; $P(q)$ and $P(q_l)$ are {\it both}
non-trivial, as in the SK model. Within mean field
theory there is no intrinsic scale for self-averaging that
grows with $L$, thereby suggesting that all observables converge
uniformily to their limits; because of that, 
we shall hereafter simply say
that the ``mean field picture'' corresponds to having
uniform convergence of observables.

\subsection{Scenarios without uniform convergence}
\label{subsect_non_uniform}
In this class of scenarios, the
$L \to \infty$ limit does not commute with the scale of
observation ($M$) going to infinity. 
In other words, there is at least one $L$-dependent scale
that affects the large $L$ limit~\cite{MarinariParisi00b}.
The simplest such case has a single scale (also called
$\ell_c$) such that the observable under
consideration (for instance the local density)
is self-averaging beyond $\ell_c$ and has a non trivial distribution below.
When $L$ is finite but large,
$P_{M,L}(q_M)$ can be approximated by considering its difference
with $P_{M}(q_M)$; it is natural to expect this difference to 
depend only on the ratio $M / \ell_c$. We can then write
\begin{equation}
\label{eq_multiscaling}
P_{M,L}(q_M) \underset{L\to\infty}{\approx}
P_M(q_M) + A_1 f_{M/ \ell_c}^1(q_M)
\end{equation}
More generally, if there were many characteristic lengths
$\ell_c^i$, we could construct such an expression 
recursively by imposing the condition $f_{0}^i = 0$; the
first term is always the pointwise limit, and the terms following
give the correction associated with passing through the
scale $\ell_c^i$. Note that once a scale $M \gg \ell_c^k$ 
is reached where $P_{M,L}$ is a single
delta function, then for observables such as density that involve
additive quantities no further change in the distribution is possible.

In their full generality, these
scenarios with multiple scales are quite complicated.
Let us expose a few possibilities in the simplest case where
there is a single $\ell_c$. That scale must grow indefinitely with
$L$, for instance as a power of $L$.
Suppose now that $P_M(q_M)$ converges at large $M$ to 
two delta function peaks centered at $q_{min}$ and
$q_{max}$, while if we take $M \gg \ell_c$ 
$P_{M,L}(q_M)$ converges to the single delta function peak
at $q$ (recall that beyond the scale $\ell_c$, $q_M$ becomes
self-averaging and thus must equal the global overlap $q$).
We have two simple candidate scenarios depending on
$q_{min}$ and $q_{max}$.

\subsubsection*{Fat sponges}
\label{subsubsect_fat}
If $q_{min}=0$ and $q_{max}=1$, then at finite scales one
never sees the surface of the cluster, while at scales larger
than $\ell_c$ the cluster is homogeneous so should ressemble
a sponge. (For this last property we assume $\ell_c \ll L$.)
We call this case the ``fat sponge'' scenario as the clusters
are just sponges whose characteristic 
mesh spacing is $\ell_c$. Indeed, up to a dilation of
the clusters, everything looks like 
the regular homogeneous scenario except that $\ell_c(L) \to \infty$ in
the infinite volume limit.

We can consider several properties of such sponges. First, since
the surface of the cluster does not grow as fast as its volume,
the standard link 
overlap converges to a delta function at $q_l = 1$, and
we have a scenario that realizes the 
TNT~\cite{KrzakalaMartin00,PalassiniYoung00a}
(trivial $q_l$, non-trivial $q$) behavior.
Second, the nature of the cluster's surface on scales smaller
than $\ell_c$ will play an important role.
The surface is unlikely to be smooth
as in the ferromagnetic case; instead, it might be continuum
fractal, having a fractal dimension
$2< d_f < 3$. But once one reaches the scale $\ell_c$, everything becomes
homogeneous as $q_M$ becomes self-averaging.

\subsubsection*{Hierarchical sponges}
\label{subsubsect_hierarchical}
Consider now the possibility that $q_{min} \ne 0$ and $q_{max} \ne 1$.
Then below the scale $\ell_c$ one is in either of two phases
differing in their density, but neither of these
two phases ressembles the ground state locally.
In fact these two phases could ressemble sponges, differing mainly
by their densities.
We will call this a ``hierarchical sponge'' scenario because
one can produce such clusters by hierarchically applying
sponge excitations to the ground state. Indeed,
start with the ground state and create a fat sponge by 
flipping a spongy cluster whose characteristic mesh 
spacing is $\ell_c$. Then if we 
take that state and flip a spongy cluster whose
characteristic mesh 
spacing is $O(1)$, we find that there are locally only
two densities and we obtain a two-step hierarchical sponge.
(Naturally this construction can be generalized to any number of 
levels.) 

\subsubsection*{Fractals}
\label{subsubsect_fractals}
Consider the fat sponge scenario but take $\ell_c$ to grow linearly
with $L$. Then there is no regime where $M \gg \ell_c$
and the cluster never becomes homogeneous. There is thus
no reason to expect it to be sponge-like (although it may
with a finite probability wrap around the lattice and have
some of the topological properties that sponges have).
The main motivation for this scenario is that it is what arises
in a disordered ferro-magnet. Its characteristic is that
$q_M$ is not self-averaging until $M \to L$ and we can expect
to get non-trivial distributions for observables such as
$q_M$ when $M$ is a finite fraction of $L$.

Overall, we see that there is a great
diversity of possible scenarios. Ideally our goal would be
to test for the uniform or not convergence of observables
as $L \to \infty$. However, from the point of view of the mean field
picture, we shall be plagued by finite size effects, whereas
from non-mean field viewpoints the results will be encouraging
but still a bit muddled. The main conclusion will be that 
$\ell_c$ grows with $L$, i.e., the clusters
coarsen as $L$ grows. This coarsening 
may stop as in the mean field scenarios, or we may be seeing evidence in
favor of the fat sponge scenario. But certainly 
$\ell_c$ does not grow linearly with $L$,
ruling out the fractal scenario having heterogeneities on the
scale $L$.

\vspace{0.5cm}
\section{Topological properties}
\label{sect_topology}

\begin{figure}
\resizebox{0.45\textwidth}{!}{
  \includegraphics{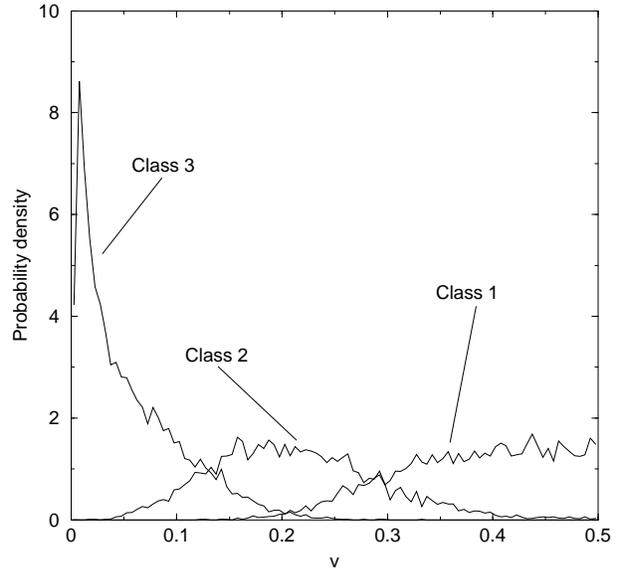}}      
\caption{Distribution of event sizes for $L=10$, decomposed
according to winding class type.}
\label{fig_dist_10}
\end{figure}

\subsection{Classification by windings}
\label{subsect_windings}
Suppose we consider all events in our data set
with $v=V/N$ in a given window; we then find
that the relative frequency of these 
events {\it decreases} as $L$ increases.
Within the droplet model, this is expected, and in fact the frequency 
should decrease to zero as $L^{-\theta}$. On the contrary,
in the mean field picture, that
frequency should go to a constant non-zero value, and
its decrease is interpreted as a finite size
effect.
Because $\theta$ is quite small,
we cannot use that data alone to
discriminate between these two pictures. Thus for each
excitation we will also look at its
topological properties as defined
from the connected cluster of spins that are flipped
in that excitation. Our
motivation is if that there are system-size excitations of
low energy, we expect them to be topologically non-trivial, and so
a joint study of topology and $v$ may allow one to extrapolate
to $L \to \infty$ cleanly. Indeed, if
the frequency of these types of events
{\it grows} rather than diminishes as $L$ increases, it
becomes plausible that they survive in the thermodynamic limit.
 
Clearly a complete topological characterization of the clusters
is not necessary. With free boundary conditions, it was
possible~\cite{KrzakalaMartin00}
to limit oneself to finding out whether the cluster
and its complement touched the different faces of the cube.
Here, since we insist on maintaining
periodic boundary conditions, 
we generalize that criterion by considering
the winding properties of the cluster in the cube. 
Given a cluster, we determine whether there are paths on that
cluster (going from site to nearest neighboring site) that wind around
in any of the three directions $(x,y,z)$ of the cubic lattice. From
this, we define three classes of clusters as follows.
A cluster (and thus the event and excitation) belongs to the 
first class if it {\it and} its complement have a non-trivial winding 
in all $3$ directions of the cube. A cluster belongs to the
third class if it has no windings at all.
Finally, the second class consists of all other events. 
We will refer to events in the first class as sponges for
obvious reasons, while events
in the third class will be referred to as droplets.

\subsection{Sponges with $O(1)$ energies}
\label{subsect_sponges}
For each excitation obtained from an instance, 
we compute its overlap $q$
with the ground state or equivalently its $v$,
and determine to which topological class it belongs. 
Figure~\ref{fig_dist_10} gives the probability density
of $v$ for each of the three classes defined.
The data represented is for $10000$ different
instances with $L=10$, and for each instance we generated
$3$ excitations.
The droplets (events in the third class)
create a peak at small $v$ while the other two classes
are responsible for the rest of the distribution. Now to understand
how these distributions evolve with $L$, we consider the total
contribution of the different classes, integrated over $v$.
From Table~\ref{tab_topology} we see that 
sponge (first class) events have an increasing frequency 
with $L$ and so can
reasonably be extrapolated to constitute a non-zero fraction in 
the large volume limit. This had also already been
the case with free boundary conditions~\cite{KrzakalaMartin00}.

\begin{table}
\begin{center}
\label{tab_topology}
\caption{Fraction of sponges and droplets}
\begin{tabular}{lll}
\hline\noalign{\smallskip}
Size & Sponges & Droplets \\
\noalign{\smallskip}\hline\noalign{\smallskip}
4 & 0.230(3) & 0.251(3) \\
5 & 0.253(3) & 0.306(3) \\
6 & 0.271(3) & 0.346(3) \\
7 & 0.284(3) & 0.374(3) \\
8 & 0.290(3) & 0.389(3) \\		
9 & 0.300(3) & 0.399(3) \\	
10 & 0.304(3) & 0.412(3) \\	
11 & 0.302(4) & 0.421(4) \\
12 & 0.310(5) & 0.423(5) \\	
\noalign{\smallskip}\hline
\end{tabular}
\end{center}
\end{table}

We feel that these data provide strong evidence that there
are system-size excitations with $O(1)$ energies.
Extrapolating to finite temperature, this leads to 
replica symmetry breaking in the 3d Edwards-Anderson model.
Furthermore, because of the constraints built into the first
class, these excitations span the whole system. From a 
direct visualisation of the clusters, we also see that their topology
is highly non-trivial: they ressemble sponges in that they
have handles everywhere on a scale of a few lattice spacings.
They are thus both space spanning (extending throughout the whole 
system) and space filling, at least on the scale of $L$.

\subsection{Are all valleys sponge-like?}
\label{subsect_non_sponges}
Given the data in Table~\ref{tab_topology}, one can ask
whether all low-energy system-size excitations asymptotically
fall into the first class. In particular, if a uniform convergence
scenario of spin glasses is correct, one expects to find only sponges 
in the thermodynamic limit as
argued in~\cite{HoudayerMartin00b}, so that the
second (intermediate) class should disappear when 
$L \to \infty$. 

\begin{figure}
\resizebox{0.45\textwidth}{!}{
  \includegraphics{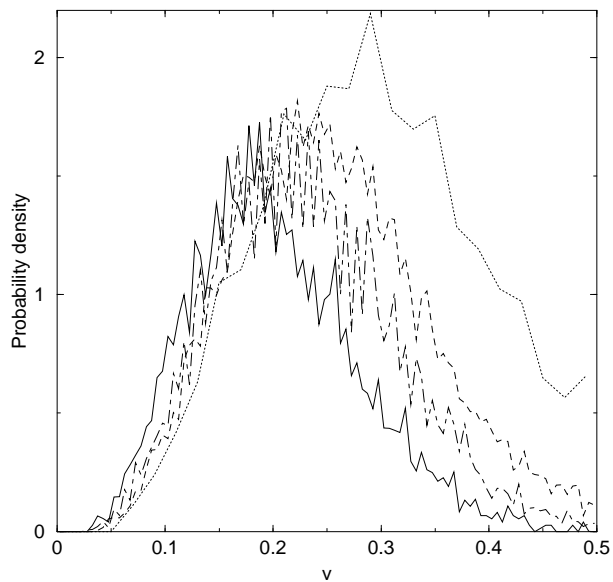}}      
\caption{Probability density of event sizes 
for $L=4,6,8,12$ from right to left; only the contribution of class 
2 events are shown.}
\label{fig_dist_class2}
\end{figure}

To see whether this is the case, we have studied the probability
density of $v$ as a function
of $L$, decomposed according to the topological classes. The 
results for class $2$ 
are displayed in figure~\ref{fig_dist_class2}. It shows 
that as $L$ increases, the curves 
shift towards smaller $v$
while the fraction of all events falling into class $2$
(given by the area under each curve)
decreases. That this fraction decreases is necessarily the case
as we saw that the fractions of the other two classes grow with
$L$. Can we extrapolate this fraction of class $2$ events
to zero as $L \to \infty$? We have performed
fits of this fraction to the functions $a + b/L + c/L^2$ and
$a + b/L^c$; both fits give reasonable 
$\chi^2$s and
lead to values for $a$ that are in reasonable agreement:
$0.18(0.01)$ and $0.17(0.01)$. 
Taken at face value, these extrapolations indicate that the
uniform scenarios are not correct. However, we 
believe that regardless of what the correct scenario is, 
finite size effects are subtle for class
$2$ events. To justify this, let us look again
at the data.

\begin{figure}
\resizebox{0.45\textwidth}{!}{
  \includegraphics{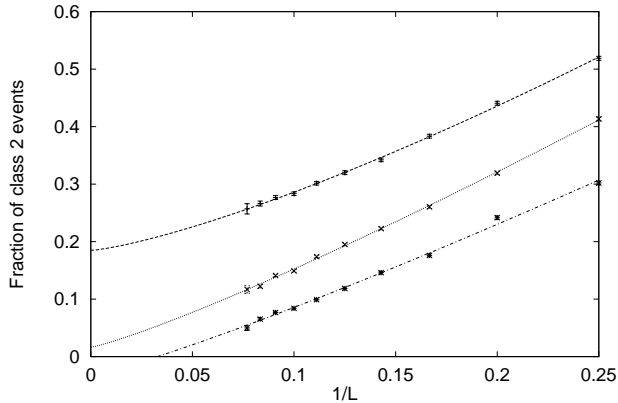}}      
\caption{Fraction of class 2 events and constant plus power fits.
From top to bottom: case of $v>0$, $v>0.2$, and $v>0.25$.}
\label{fig_non_vanish}
\end{figure}

In Figure~\ref{fig_non_vanish} we show the fraction of 
class $2$ events as a function of $1/L$ (top curve). 
Because of the positive curvature of this data, it seems 
likely that this class survives in the thermodynamic limit, 
confirming the conclusion drawn from the fits.
But as argued in~\cite{HoudayerMartin00b}, it is difficult to 
create sponges in finite dimensions when $L$ is smaller than
a typical ``cohesion'' length $\ell_c$ that may be substantially 
bigger than one lattice spacing, especially if the
dimension is not large. This effect should be much more
severe for small $v$; indeed in that limit, 
creating a spongy yet connected cluster requires the sponge 
to be very ``thin'',
most of its sites being at its surface. The energy of such a sponge
is likely to be high, and so our procedure for generating large scale
excitations should not pick them up.
We have explored the validity of this point of view by considering the
fraction of class $2$ events subject to a further selection
in $v$. The three curves of Figure~\ref{fig_non_vanish}
are associated with the events satisfying $v > 0$, 
$v > 0.2$, and $v > 0.25$ (from top to bottom). We see that 
the fraction of these last events {\it cannot} be extrapolated to
positive values, as confirmed by detailed fits. In fact, 
when we look
at Figure~\ref{fig_dist_class2}, we can now reasonably assume 
that the curves all go to zero when $L$ grows, but the
smaller $v$ is, the later the
asymptotic behavior sets in. More convincing evidence of this
is beyond the scope of our data, but we hope to have made
the case that our results are not incompatible with class $2$ events
disappearing altogether in the large $L$ limit. 

In conclusion, great care is needed when performing extrapolations
to large $L$ because finite size effects are subtle
when $v$ is small. Therefore, in what follows, 
we will focus exclusively
on the region near $v=1/2$ ($q=0$). If we are to provide any
evidence that there is a growing length scale, 
it is only at $q \approx 0$ that such evidence can be solid.

\section{Geometrical properties}
\label{sect_geometry}
We now move on to geometrical properties 
of low-energy system-size excitations: 
link overlaps, correlation functions, window overlaps, etc...
From these measurements, we hope to test whether or not 
there is a growing intrinsic scale $\ell_c$, and to thereby
weight the balance in favor
of some of the scenarios. Since spin glasses have resisted
simple solutions, we do not expect the reader to
find that any scenario comes out a clear winner but
we do hope to convince her or him that 
the clusters coarsen slignificantly as $L$ grows. Whether or not this
coarsening continues to infinite sizes may be of fundamental importance;
however, if it turns out that the growth is very slow, it should be
of limited experimental relevance.

\subsection{Mean link overlaps}
\label{subsect_meanql}
We begin by considering
the link overlap between the ground state and an excited state:
\begin{equation}
q_l = {\frac{1}{3 L^3}} \sum_{i, {\hat {\mu} } }
S^0(i) S^1(i)  S^0(i+{\hat{\mu}}) S^1(i+{\hat{\mu}})
\end{equation}
We will take the surface $S$
of the corresponding cluster to be the number of links that connect
the cluster to its complement. Then we have
$q_l=1-2 S / 3 L^3$. One of the main issues in this section
is whether $q_l \to 1$ at large $L$, that is we 
want to know whether the system-size clusters have a surface 
to volume ratio that goes to zero at large $L$.

\begin{figure}
\resizebox{0.45\textwidth}{!}{
  \includegraphics{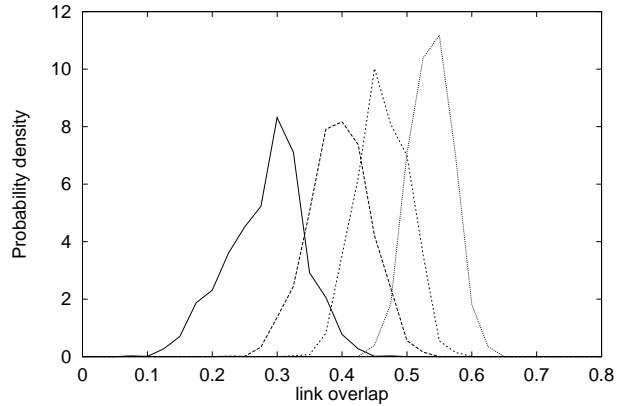}}      
\caption{Link overlap distributions for $L=4, 6, 8, 12$. 
The data are for all events with $0 \le q \le 0.1$.}
\label{fig_pofql}
\end{figure}

In the SK model, $q_l$ is a deterministic function of
the spin overlap $q$ in the thermodynamic limit: $q_l = q^2$. An
analogous deterministic relation will be satisfied in any scenario
that is homogeneous on the length scale $L$. We can see whether
this is the case in our system by considering 
the probability distribution
of $q_l$ at fixed $q$; Figure~\ref{fig_pofql}
shows this distribution for different $L$ values when
$0<q<0.1$. 
(Note that for such small values of $q$, essentially all the 
excitations are sponge-like, i.e., fall into class $1$ except for the smallest
$L$s, and so a nearly identical plot is obtained
if one restricts oneself to class $1$ events.) 
Clearly the distributions are becoming more narrow
with increasing $L$ while drifting to the right.
So we will look at both the mean and variance of these distributions.

In the uniform convergence scenarios
the mean of these curves should settle at a $q_l < 1$ as
$L \to \infty$ and so the drift is a finite size 
effect. On the contrary, in the fat sponge and fractal scenarios, 
the limiting $q_l$ is equal to $1$ and the drift is the signal that
$\ell_c$ is increasing with $L$. (In the hierarchical sponge
scenario, $q_l$ should again converge to a value strictly 
smaller than $1$.)

In geometrical terms, having $q_l \to 1$ means that
the surface to volume ratios of the clusters go to
zero at large $L$. Such a property was referred to
as TNT in~\cite{KrzakalaMartin00} as the link
overlap $q_l$ then has a trivial probability 
distribution~\cite{PalassiniYoung00a}
(it is infinitely narrow and localized at $q_l=1$) while
the distribution of the spin overlap $q$ remains non-trivial. 

We can try to discriminate between 
such different scenarios by performing fits, but it is necessary to
parametrize the dependence of the mean on $L$ for each
scenario.
Suppose that the asymptotic behavior of a quantity
is $L^{\alpha}$; then we postulate that the finite size effects
are multiplicative, with a finite size correction given by a
function $f(1/L)$ where the argument is simply the ratio of
the two scales of the problem, i.e., the lattice spacing and the lattice
size. It is natural to assume that $f$ 
can be Taylor expanded
for small arguments, so we shall parametrize all our finite
size effects by polynomials in $1/L$.

\begin{figure}
\resizebox{0.45\textwidth}{!}{
  \includegraphics{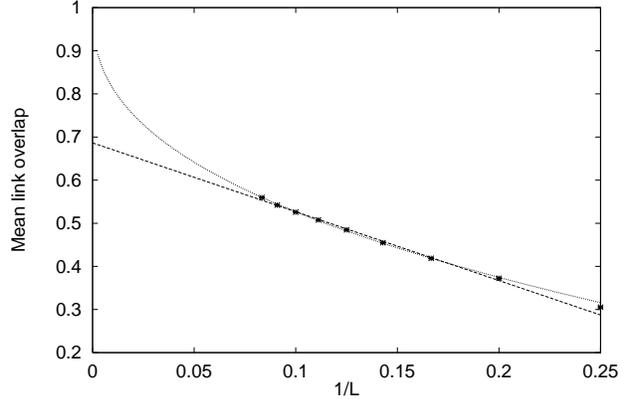}}      
\caption{Mean link overlap for excitations 
with $0<q<0.1$. Shown are fits for constant and
power law asymptotics.}
\label{fig_meanql}
\end{figure}

In TNT scenarios, one expects $\langle q_l \rangle -1$ to
go to zero as an inverse power of $L$, whereas in the 
scenarios with uniform convergence $\langle q_l \rangle -1$ goes
to a strictly positive constant.
We have performed fits for these two types of scaling behavior. 
(We have also considered the possibility that
$\langle q_l \rangle -1$ vanishes as an inverse
power of the logarithm of $L$, but the results are not
convincing.)
Consider first the fits having {\it two} free 
parameters. There is a linear
fit in $1/L$, and a pure power fit, 
$\langle q_l \rangle -1 = a / L^{\gamma}$, as shown in
figure~\ref{fig_meanql}. We do not include
the $L=4$ data when fitting, and obtain 
$\chi^2$s of $67$ and $19.5$ for the two fits.
(If instead we do leave the point at $L=4$, the $\chi^2$s are
increased to $149$ and $49$.) It seems plausible
that we need to allow for more parameters and finite size corrections.
So we allow now for three fitting parameters.
The new $\chi^2$s are $7.5$ and $6.3$.
This same pattern is repeated when one goes on to
$4$ free parameters. Certainly the case of a power
law in $L$
leads to the best fits and so is favored by the data.
However, this advantage is only significant for the two parameter
fits. (Note that when repeating all the fits after restricting
the events to belong the sponge class, we find similar results.) 
Since it is quite possible that our parametrizations
do not capture correctly the finite size effects, we cannot
rule out the scenarios where $\langle q_l \rangle$
does not go to $1$. 
Even so, it is worrisome that the asymptotic $\langle q_l \rangle$
values predicted by such scenarios drift with the number
of parameters in the fits: we obtain 
$\langle q_l (L = \infty ) \rangle = .68$, $.72$, and
$.75$ for the fits with $2$, $3$, and $4$ parameters.

To have a completely convincing test, we would need to 
have a range in $L$ where only one of the fits gave
$\chi^2$s that were close to the theoretically
expected value. As can be seen in the
figure, such a situation may not be so far: it might be enough
to go to sizes up to $L=16$ to finally obtain a 
clear cut discrimination of the different scenarios. In the mean
time, we shall simply say that the TNT scenarios
are slightly preferred.

Regardless of the scenario favored by the reader, perhaps the
most striking point made here is that
$\langle q_l \rangle$ undeniably increases for the currently 
accessible values of $L$, and that this is most 
simply interpreted by having $\ell_c$ growing
with $L$. In models without geometry
$\ell_c$ is small (for instance in the SK it is $1$) whereas it seems
to be at least $4$ here even if we stick to the
uniform convergence scenarios and identify $\ell_c$ with
$1/(1-q_l)$.

\subsection{Variance of link overlaps}
\label{subsect_varql}
We come back now to the way the distribution of
$q_l$ becomes narrow. If the lattice spacing is the only 
relevant scale as in the homogeneous sponge
scenario, we can expect $P(q_l)$ to converge
to a delta function following the central limit theorem
scaling law:
\begin{equation}
\label{eq_sigmaQl}
\langle (q_l - \langle q_l \rangle)^2 \rangle \approx C L^{-3}
\end{equation}
However the data very clearly shows that 
$\langle (q_l - \langle q_l \rangle)^2 \rangle \times L^3$ grows
tremendously with $L$ and we see no way this
can be compatible with finite size corrections to a constant
asymptotic value. On the contrary, a fit to a power law looks
sensible and we find that 
$\langle (q_l - \langle q_l \rangle)^2 \rangle \times L^{1.1}$
has no trend with $L$. This value for the exponent of the scaling
of the variance of $q_l$ is not very different from the value found
by Marinari and Parisi~\cite{MarinariParisi00c} who extracted
excitations in a very different way. This scaling
suggests to us that there
is an underlying length $\ell_c$ that diverges with $L$
and that the density of the 
cluster's interface fluctuates on scales much larger than the lattice
spacing. We also have measured the kurtosis of the distribution; we find that
it does not decrease with $L$, showing explicitly that there is no
central limit behavior.
Overall, we feel that these results rule out the homogeneous sponge
scenario for which the
central limit theorem should hold. However we have nothing to say
concerning the heterogeneous sponge scenario which can very well violate
the central limit theorem.

\subsection{Spin overlap correlation functions }
\label{subsect_Gofr}
Now we focus on the two-point correlation function 
defined by:
\begin{equation}
\label{eq_Gofr}
G(L,r) = \langle S^0(i) S^1(i)  S^0(i+r) S^1(i+r) \rangle
\end{equation}
\begin{figure}
\resizebox{0.45\textwidth}{!}{
  \includegraphics{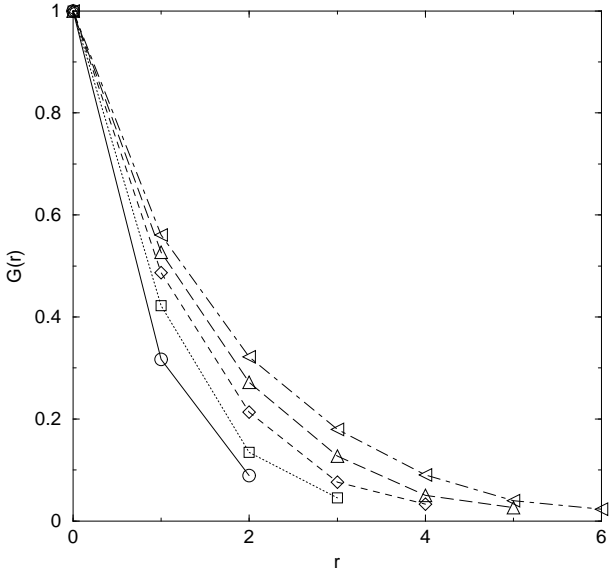}}    
\caption{Spin overlap correlation function
at $0<q<0.1$ for $L=4,6,8,12$ (bottom to top).}
\label{fig_GdeR}
\end{figure}
We present in figure~\ref{fig_GdeR} this correlation function,
as usual for events with $0<q<0.1$. We see
again that finite size effects are very severe,
making it quite difficult to guess what is the
limiting curve when $L \to \infty$. In the regular scenarios, one
expects $G(r,L)$ to converge to a limiting curve with for instance
$1/L$ corrections, just as we saw for the mean link overlaps.
For the TNT scenarios, the limiting curve is
$G(r, L\to \infty) = 1$, and we can follow the discussion of
finite size effects for $\langle q_l \rangle$ to motivate 
those of $G(r,L)$. (Note in particular that
$G(1,L) = q_l$.) Since $G(r,L)$ probes properties at distance $r$,
we will postulate in the fat sponge scenario that the
finite size effects can be parametrized through
\begin{equation}
\label{eq_Gofr_lc}
G(r,L) = 1 - A(r) L^{-\delta}
(1 + {\frac{B_1(r)}{L}} + \cdots )
\end{equation}
in the region $r \ll \ell_c$, while $G(r,L)$ should
be constant and equal to $q^2$ for
$r \gg \ell_c$. 

We have performed these fits.
Since it was possible to fit reasonably well $G(1,L)$ to a constant different
from $1$ plus $1/L$ corrections, (cf. the previous section),
it will be no surprise to find out that this remains
true of $G(2,L)$. The difference between the $r=1$ and
$r=2$ cases is that now the quality of the fits
with $G(r, L=\infty)$ equal to $1$ or less than $1$ are
similar, no scenario is favored. We have also tried to see whether
a simple scaling law of the type $G(r,l) = G(r/L^{\delta})$ might
hold, but given the small useful range in $r$, no solid conclusion
can be drawn. However, there should be a large usable range
in $r$ if we assume 
$G(r,l) = G(r/L)$ as in the fractal scenario. When we try that rescaling
the data do not superpose at all:
the large $L$ curves that were above are now
far below the others. There seems no way to avoid the conclusion that
the correlation function data unambiguously 
rule out fractal scenario.

\subsection{Window overlaps}
\label{subsect_windows}
\begin{figure}
\resizebox{0.45\textwidth}{!}{
  \includegraphics{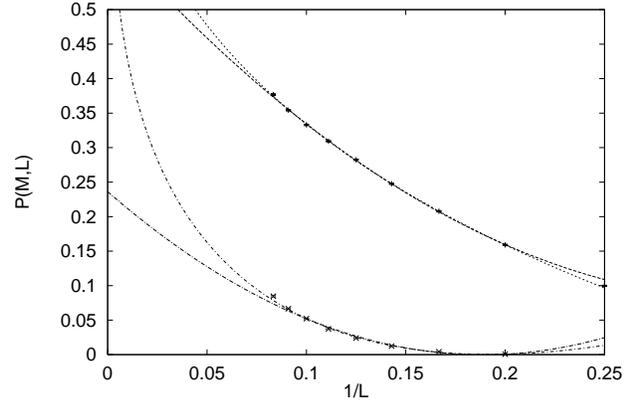}}    
\caption{Probability and fits that an $M^3$ box does not cross the surface 
of a cluster for $0<q<0.1$ events. Top: $M=2$; bottom: $M=3$.}
\label{fig_box}
\end{figure}

Now consider the probability $P(M,L)$ that
a cubic box of size $M$ is entirely contained either in
the cluster or in its complement. $P(M,L)$ has been 
studied in previous work, in particular
by Palassini and Young~\cite{PalassiniYoung99a}.
Figure~\ref{fig_box} gives these probabilities for our excitations
at $M=2$ and $3$ for the window $0<q<0.1$.
In any TNT scenario and in particular in the
fat sponge scenario, $P(M,L) \to 1$ as $L \to \infty$
at any fixed $M$. We have performed the fits to this quantity
in the same way as previously.
The mean field picture is associated with polynomial fits in $1/L$,
while in the fat sponge scenario we take
\begin{equation}
P(M,L) = 1 - A(M) L^{-\delta} ( 1 + {\frac{B_1(M)}{L}} + \cdots ) 
\end{equation}
For $M=2$, we find the following $\chi^2$s for these two
pictures:
$311$ and $10.6$ for two free parameters,
$14.4$ and $7.5$ for three free parameters, and
$8.4$ and $7.2$ for four free parameters. (The same pattern 
transpires if we include the $L=4$ data.) So here we seem to find
more convincing evidence that the fat sponge scenario is significantly
favored over the uniform convergence scenarios. The same study
at $M=3$ leads to the same conclusion, i.e., that the fat sponge scenario
is favored.

\subsection{Tube observables}
\label{subsect_tubes}
In the context of the fat sponge scenario with a length scale
$\ell_c$, we may expect $P(M,L)$ to be a function only of
the reduced variable $M/\ell_c(L)$. Unfortunately, the usable
range in $M$ is small (the data at $M=4$ is not 
exploitable), and so the scaling cannot be reliably tested. This leads us to
consider other observables that may have a larger useful range.
\begin{figure}
\resizebox{0.30\textwidth}{!}{
  \includegraphics{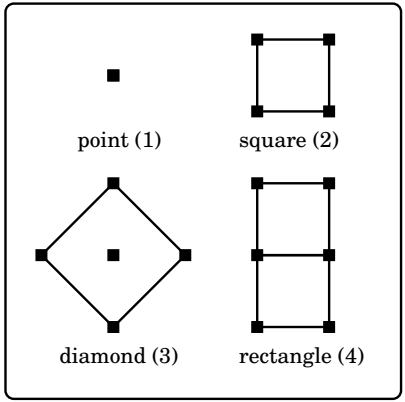}}
\caption{The cross-sections of our four types of tubes.}
\label{fig_Tube_Type}
\end{figure}

Rather than increase the size of the window in all three directions,
we increase it in just one direction; because of this, we refer to
these windows as ``tubes''. We have investigated four different
types of cross-sections for these tubes, as displayed
in figure~\ref{fig_Tube_Type}. Essentially, we use these objects
to probe the size of holes in the clusters as follows. Given
a tube type and a lattice site, 
we ask for the length of the longest tube 
along the $x$ axis that passes through this site and
is entirely contained in the
cluster (or in its complement). We compute this length for each
site of the lattice and for the four different tube types. Then we
extract the mean length found, averaged over 
lattice sites and over the disorder.

\begin{figure}
\begin{center}
\resizebox{0.45\textwidth}{!}{
  \includegraphics{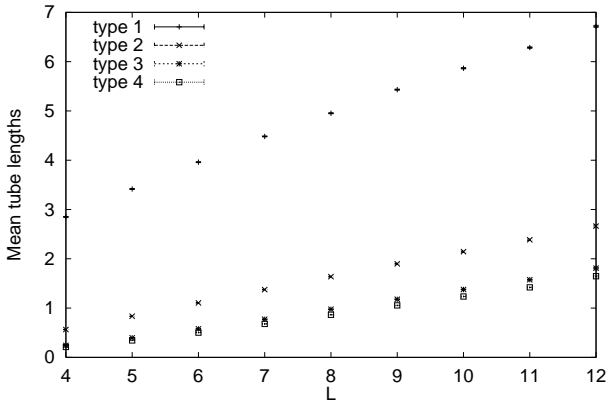}}
\caption{Mean tube length for events
with $0<q<0.1$ as a function of $L$.}
\label{fig_tubeLengths}
\end{center}
\end{figure}

\begin{figure}
\resizebox{0.45\textwidth}{!}{
  \includegraphics{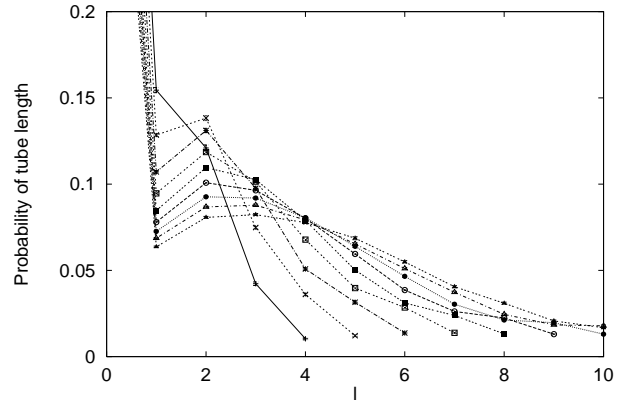}}      
\caption{Histogram of type $2$ tube lengths for
events with $0<q<0.1$, for $4 \le L \le 12$.}
\label{fig_tube1Dist}
\end{figure}

In figure ~\ref{fig_tubeLengths} we see 
that the mean tube length
increases very clearly as $L$ increases, even for events with
$0<q<0.1$.
Within the fat sponge scenario, this is simply the reflection of
the growth of $\ell_c$ with $L$.
Is such a behavior compatible with the heterogeneous
sponge scenario (having uniform convergence)?
If the distribution of hole sizes is broad in heterogeneous sponges,
the first moment may no longer be finite, leading to a divergence in
the mean tube length. It is thus necessary to look directly at
the probability distribution of these lengths. 
In Figure~\ref{fig_tube1Dist} we show the distribution of the 
tube lengths for tubes of type $2$. (Tubes of type $1$
show a much more severe drift with $L$, and the other tube types
are quite similar to the one displayed.)
We see that as $L$ increases,
the curves shift to the right while the values at small
$l$ decrease. In the uniform convergence scenarios, 
these distributions have a limiting shape, while in
the fat sponge scenario for instance, the values at any finite $l$ will go to 
zero. We have fitted the first three values ($l=0, 1, 2$) in
the figure 
as we have previously, i.e., to a finite or null asymptote, 
using the data with $L > 5$.
The fits at $l=0$ have the same kind of behavior we saw before,
i.e, the best fit is the extrapolation to zero, but
the pure $1/L$ fit should not be excluded. (For two parameter fits,
the $\chi^2$s are $27$ for a non-zero extrapolation and
$8.6$ for the zero extrapolation; for the $3$ and $4$ parameter
fits the zero extrapolations remain the best but are only marginally
better than the ones for non-zero extrapolation.)
For $l=1$ and $2$, there is no trend, but the extrapolated values
are always very small, being below $0.03$ and often ten times smaller
than that. This result could have been guessed at from looking
at Figure~\ref{fig_tube1Dist}. Naturally, given the form of these
histograms, it is much too difficult to extrapolate the data
for still larger values of $l$, but nevertheless our conclusion is
that the data do not support a point-wise convergence to a non-trivial
distribution. At face value, this means that the uniform convergence
scenarios are incorrect.

We can also look at the tail of the distribution. If $L \gg \ell_c$,
we can expect the tail of the histogram to fall to zero rather rapidly
with $L$. In practice though, we find that the histogram at
$l=L$ grows with $L$ for the tube types $3$ and $4$ for all values
explored, while it first grows then decreases very slowly 
for type $2$ tubes. At the very least, this is totally unexpected
within the uniform convergence scenarios. Because of this, we have
investigated the probability that a cluster has somewhere a
tube of maximum length (that is of length $L$, the tube
spanning the whole system along the $x$ direction).
We find that for tubes of the first type, $99\%$ of the 
instances have at least one spanning tube, and that this ratio does not
seem to depend on $L$. For the other tube types, we have collected
the ratios in table~\ref{tab_maxTube}. One sees that the ratios
all increase with $L$. If we take into account the data for
tubes of the first type, we are lead to extrapolate these
ratios to finite quantities, distinct from $0$ and $1$. This goes 
strongly against the idea that the clusters are homogeneous beyond
$\ell_c \ll L$. Indeed, in such a picture, the probability of having
a tube span the whole system decreases exponentially. Here we need
the decrease to be power-like so it can be compensated
by the $L^2$ places where the tube can be located.
Thus we are driven to a rather heterogeneous picture
(in the sense discussed in Section~\ref{subsubsect_inhomogenous})
where large
fluctuations on scales larger than $\ell_c$ are possible, being suppressed
by powers of $L$ rather than exponentially in $L/\ell_c$.

\begin{table}
\begin{center}
\label{tab_maxTube}
\caption{Fraction of excitations having a spanning tube}
\begin{tabular}{llll}
\hline\noalign{\smallskip}
L & Tube 2 & Tube 3 & Tube 4 \\
\noalign{\smallskip}\hline\noalign{\smallskip}
4 & 0.137(7) & 0.006(2) & 0.008(2) \\
5 & 0.22(1) &  0.028(4) & 0.020(3) \\
6 & 0.32(1) & 0.080(6) & 0.043(5) \\
7 & 0.38(1) & 0.119(8) & 0.080(6) \\
8 & 0.41(1) & 0.15(1)  & 0.11(1) \\		
9 & 0.46(1) & 0.18(1)  & 0.13(1) \\	
10 & 0.47(1) & 0.22(1) & 0.17(1) \\	
11 & 0.48(1) & 0.25(1) & 0.20(1) \\
12 & 0.51(2) & 0.28(2) & 0.24(2) \\	
\noalign{\smallskip}\hline
\end{tabular}
\end{center}
\end{table}

\section{Conclusions}
\label{sect_conclusions}
The only truely
compelling property that emerges from this study is that
the clusters coarsen as $L$ grows. Naturally, this can
be interpreted within the uniform scenarios as being a finite
size effect; the issue is whether this is believable given the
size of this effect. In essentially all the measurements
we have made, the fat sponge type picture was favored 
over the homogeneous sponge scenario
by a detailed $\chi^2$ analysis. Furthermore
as we saw from the tube observables, heterogeneities are 
large, so a scenario that is homogeneous on the scale of the
lattice spacing seems ruled out. (We also saw for instance
from the measurements of $G(r,L)$ that the scenarios where
$L$ is the only relevant scale are not supported by the data.)
So our conclusion is mainly that the scenarios at
the end of the spectrum (homogenous regular sponge
and fractal or droplet-like model) are no longer compelling.
We are thus left with uniform convergence scenarios
having large heterogeneities, and scenarios with
at least one scale $\ell_c$ growing with $L$. In the
first case, we have large finite size corrections because
of the heterogenities, while in the second we can realize the
TNT scenario where $q_l$ has a trivial distribution in
the thermodynamic limit. 

The difficulty of the problem should be all the more evident
from the focus of our study: essentially all of our measurements
were for the region $q \approx 0$. Obviously there is room
for much more heterogeneities when $q$ is larger. Nevertheless
we hope to have shed new light on the problem of replica
symmetry breaking in finite dimensional systems. And, as we have
pointed out before, a convincing test of the major
scenarios may not be so far off: we believe it is enough
to go to slightly larger systems before the tests will become
clear-cut to a vast majority.

\vspace{1cm}

\centerline{{\bf Acknowledgments}}
We thank J.-P. Bouchaud and M. M\'ezard for very stimulating
discussions and for their continuous encouragement.
J.H acknowledges financial support from the Max Planck Institut,
F.K. a fellowship from the
MENRT, and O.C.M. support from the Institut Universitaire de
France. The LPTMS is an Unit\'e de Recherche de
l'Universit\'e Paris~XI associ\'ee au CNRS.

\bibliographystyle{plain}
\bibliography{../../../Bib/references}

\addcontentsline{toc}{chapter}{\protect\bibname}
\begin{thebibliography}{10}

\bibitem{BrayMoore86}
A.~J. Bray and M.~A. Moore.
\newblock Scaling theory of the ordered phase of spin-glasses.
\newblock In J.~L. van Hemmen and I.~Morgenstern, editors, {\em Heidelberg
  Colloquium on Glassy Dynamics}, volume 275 of {\em Lecture Notes in Physics},
  pages 121--153, {B}erlin, 1986. {S}pringer.

\bibitem{EdwardsAnderson75}
S.~F. Edwards and P.~W. Anderson.
\newblock Theory of spin-glasses.
\newblock {\em J. Phys. F}, 5:965--974, 1975.

\bibitem{FisherHuse88}
D.~S. Fisher and D.~A. Huse.
\newblock Equilibrium behavior of the spin-glass ordered phase.
\newblock {\em Phys. Rev. B}, 38:386--411, 1988.

\bibitem{HoudayerMartin99b}
J.~Houdayer and O.~C. Martin.
\newblock Renormalization for discrete optimization.
\newblock {\em Phys. Rev. Lett.}, 83:1030--1033, 1999.
\newblock cond-mat/9901276.

\bibitem{HoudayerMartin00b}
J.~Houdayer and O.~C. Martin.
\newblock A geometrical picture for finite dimensional spin glasses.
\newblock {\em Euro. Phys. Lett.}, 49:794--800, 2000.
\newblock cond-mat/9909203.

\bibitem{KrzakalaMartin00}
F.~Krzakala and O.~C. Martin.
\newblock Spin and link overlaps in three-dimensional spin glasses.
\newblock {\em Phys. Rev. Lett.}, 85:3013--3016, 2000.
\newblock Cond-mat/0002055.

\bibitem{MarinariParisi00b}
E.~Marinari and G.~Parisi.
\newblock Effects of changing the boundary conditions on the ground state of
  {I}sing spin glasses.
\newblock {\em Phys. Rev. B}, 62:11677, 2000.
\newblock cond-mat/0005047.

\bibitem{MarinariParisi00c}
E.~Marinari and G.~Parisi.
\newblock On the effects of a bulk perturbation on the ground state of 3d
  {I}sing spin glasses.
\newblock 2000.
\newblock cond-mat/0007493.

\bibitem{MarinariParisi99b}
E.~Marinari, G.~Parisi, F.~Ricci-Tersenghi, J.J. {Ruiz-Lorenzo}, and
  F.~Zuliani.
\newblock Replica symmetry breaking in short range spin glasses: A review of
  the theoretical foundations and of the numerical evidence.
\newblock {\em J. Stat. Phys.}, 98:973, 2000.
\newblock cond-mat/9906076.

\bibitem{MezardParisi87b}
M.~M{\'e}zard, G.~Parisi, and M.~A. Virasoro.
\newblock {\em Spin-Glass Theory and Beyond}, volume~9 of {\em Lecture Notes in
  Physics}.
\newblock World Scientific, Singapore, 1987.

\bibitem{PalassiniYoung99a}
M.~Palassini and A.~P. Young.
\newblock Triviality of the ground state structure in {I}sing spin glasses.
\newblock {\em Phys. Rev. Lett.}, 83:5126--5129, 1999.
\newblock cond-mat/9906323.

\bibitem{PalassiniYoung00a}
M.~Palassini and A.~P. Young.
\newblock Nature of the spin glass state.
\newblock {\em Phys. Rev. Lett.}, 85:3017, 2000.
\newblock cond-mat/0002134.

\bibitem{Young98}
A.~P. Young, editor.
\newblock {\em Spin Glasses and Random Fields}.
\newblock World Scientific, Singapore, 1998.

\end{thebibliography}

\end{document}